\documentclass[12pt]{iopart}
\usepackage{graphicx,bm,url,color,cite}
\newcommand{\dd}{\,{\rm d}}
\newcommand{\days}{\,{\rm days}}
\newcommand{\const}{{\rm const}}
\newcommand{\xx}{\bm{x}}
\newcommand{\Tab}[1]{Table~\ref{#1}}
\newcommand{\Fig}[1]{Figure~\ref{#1}}
\newcommand{\Figs}[2]{Figures~\ref{#1} and \ref{#2}}
\newcommand{\Eqs}[2]{Equations~(\ref{#1}) and~(\ref{#2})}
\newcommand{\bra}[1]{\langle #1\rangle}

\begin{document}

\title[Quadratic growth during the COVID-19 pandemic]{Quadratic growth during the COVID-19 pandemic: merging hotspots and reinfections}

\author{Axel Brandenburg$^{1,2,3,4}$}

\address{
$^1$Nordita, KTH Royal Institute of Technology and Stockholm University, Hannes Alfv\'ens v\"ag 12, SE-10691 Stockholm, Sweden \\
$^2$The Oskar Klein Centre, Department of Astronomy, Stockholm University, AlbaNova, SE-10691 Stockholm, Sweden\\
$^3$McWilliams Center for Cosmology and Department of Physics, Carnegie Mellon University, 5000 Forbes Ave, Pittsburgh, PA 15213, USA\\
$^4$School of Natural Sciences and Medicine, Ilia State University, 3-5 Cholokashvili Avenue, 0194 Tbilisi, Georgia
}
\ead{brandenb@nordita.org}
\vspace{10pt}
\begin{indented}
\item[]\today
\end{indented}

\begin{abstract}
The existence of an exponential growth phase during early stages of a
pandemic is often taken for granted.
However, for the 2019 novel coronavirus epidemic, the early exponential phase
lasted only for about six days, while the quadratic growth prevailed for forty
days until it spread to other countries and continued, again quadratically,
but with a larger coefficient.
Here we show that this rapid phase is followed by a subsequent slow-down
where the coefficient is reduced to almost the original value at the
outbreak.
This can be explained by the merging of previously disconnected sites
that occurred after the disease jumped (nonlocally) to a relatively small
number of separated sites.
Subsequent variations in the slope with continued growth can qualitatively
be explained as a result of reinfections and changes in their rate.
We demonstrate that the observed behavior can be described by a standard
epidemiological model with spatial extent and reinfections included.
Time-dependent changes in the spatial diffusion coefficient can also
model corresponding variations in the slope.
\end{abstract}

\vspace{2pc}

\submitto{\jpa}

%
\ioptwocol

\section{Introduction}

Soon after the news about the 2019 novel coronavirus epidemic emerged,
people in Europe followed the increasing case numbers with concern
\cite{Backer+20, Zhou+20, Singer+20, Wu+20b, Britton+20, Prasse+20,
Chen+20, Wu+20, Britton20, Wang+20, Tang+20, Roosa+20}.
The first deaths occurred on January 20, and, for a long time, the
ratio of the number of deaths to that of cases was around 0.02.
Even today (December 2022), with $6.7\times10^6$ deaths and
$6.6\times10^8$ cases worldwide, the ratio is still about 0.01.

\begin{figure*}[t]
\begin{center}
\includegraphics[width=\textwidth]{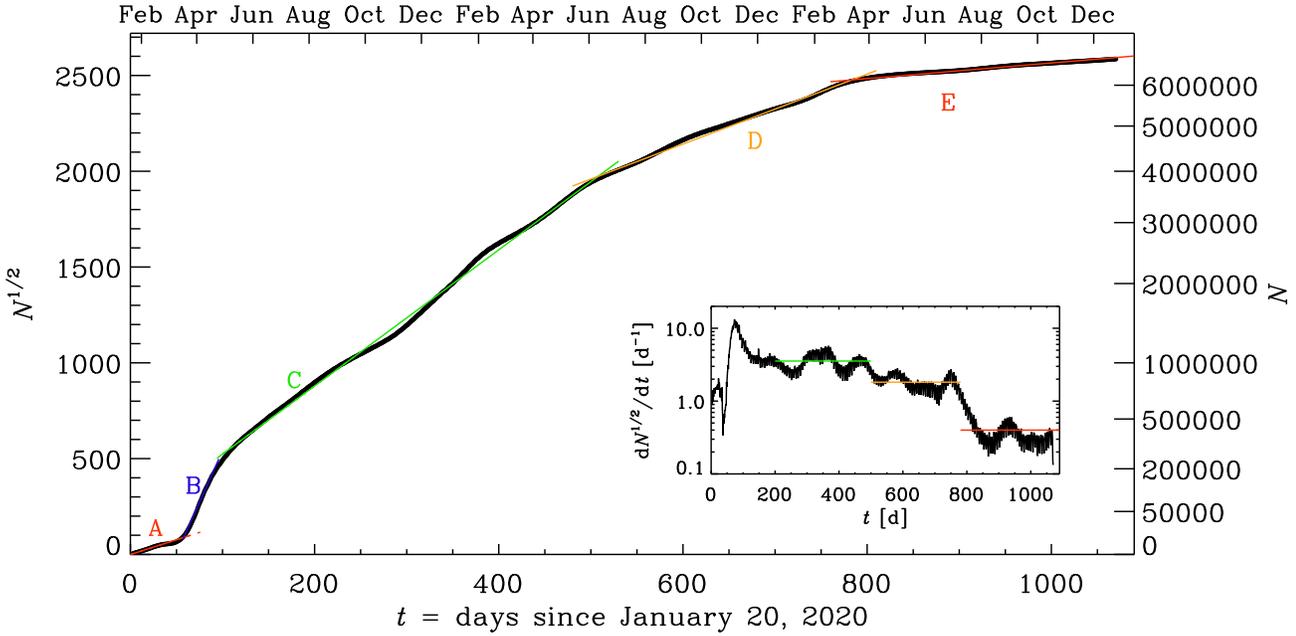}
\end{center}
\caption{Square root of the number $N$ of deaths, which is regarded as a
proxy of the number of infected that is more reliable than the reported
number of SARS-CoV-2.
Note the piecewise linear growth in $N^{1/2}$, corresponding to a
piecewise quadratic growth.
The line segments A--E are described in the text.
}\label{p3extn}
\end{figure*}

It soon became clear that the number of cases increased subexponentially
\cite{Bra20, Ziff+Ziff20, Maier+Brockmann20, Bodova+Kollar20,
Radicchi+Bianconi20, Blanco+21, Triambak+21, Rast22}.
This was called peripheral growth \cite{Bra20}, which means that the rate
of increase of the number of cases or deaths is proportion to the length
$\ell$ of the periphery of a patch on a map containing the population
with the disease.
If it is just a circular patch of radius $r$, the circumference is
$\ell=2\pi r$ and the number of cases is $N_i=n_i\pi r^2$, where $n_i$ is
the density of cases ($i={\rm C}$) or deaths ($i={\rm D}$) per unit area.
The rate of increase of $N_i$ is then
\begin{equation}
\frac{\dd N_i}{\dd t}=\alpha\ell=2\alpha (\pi N_i/n_i)^{1/2},
\quad i={\rm C}, {\rm D}
\label{rate}
\end{equation}
for ``cases'' and ``deaths'' with the solution
\begin{equation}
N_i^{1/2}=N_{i0}^{1/2}+(t-t_0)/\tau,
\label{soltn}
\end{equation}
where $N_{i0}=N_i(t_0)$ is the initial condition at $t=t_0$
and $1/\tau=\alpha (\pi/n_i)^{1/2}$ is the slope.
Thus, we expect a quadratic growth where $N_i^{1/2}$ versus
$t$ increases linearly or piecewise linearly with $t$.
This was clearly seen in the original study in Ref.~\cite{Bra20}.
Subsequent work confirmed the existence of algebraic growth, although
the exponent was sometimes found to deviate from 2.
This could be related to growth on a fractal \cite{Ziff+Ziff20}.
In \Fig{p3extn}, we present an updated plot of the square of the number
of deaths,\footnote{\url{http://www.worldometers.info/coronavirus/}}
$N_{\rm D}^{1/2}$, versus $t$.
We identify five different slopes, A--E, whose values and their reciprocal
values are given in \Tab{TAB}.

Of particular importance is the fact that, at some point, the quadratic
growth sped up; see \Fig{p3extn} at $t\approx50\days$ after January 20,
2020, or Figure~2 of Ref.~\cite{Bra20}.
This was possible to model through the emergence of multiple nucleation
sites from where the disease spread.
This meant that the coefficient $\alpha$ should be replaced by
$\alpha\to\alpha_1+\alpha_2+...+\alpha_M$, depending on the number
$M$ of sites, increasing thereby the slope of $N_i^{1/2}$.

It was already anticipated in Ref.~\cite{DATA} that
the subsequent decrease of the growth of the number of cases and deaths
could be associated with the merging of independent sites from which
separated fronts continue to expand after an initial period of quadratic
increase in the number of cases or deaths.
The details of this process were not, however, explored in detail.
This is the purpose of the present paper.
We begin by discussing the spatially extended version of the standard
$SIR$ model \cite{KMK27}, where $S$ stands for the number of susceptible
individuals, $I$ for the number of infectious individuals, and $R$
for the number of recovered, deceased, or immune individuals.

\section{The spatially extended $SIR$ model}

In epidemics, the $SIR$ model and its extensions are an important corner
stone in the theory of epidemics.
The extension to including a diffusion operator, i.e., $\kappa\nabla^2$,
where $\kappa$ is the diffusivity, is important when complete mixing
among the population can no longer be assumed.
Their inclusion has dramatic consequences for the evolution of the total
number of cases or deaths.
In the following, we discuss the consequences in detail.
We begin by outlining the essence of the $SIR$ model

\begin{table}[t!]\caption{
Parameters of the five intervals of \Fig{p3extn}.
}\vspace{12pt}\centerline{\begin{tabular}{ccc}
Interval & $1/\tau$ [$\dd^{-1}$] & $\tau$ [$\dd$] \\
\hline
A &  1.54 & 0.65 \\
B & 10.7 & 0.09 \\
C &  3.55 & 0.28 \\
D &  1.82 & 0.55 \\
E &  0.41 & 2.5 \\
\label{TAB}\end{tabular}}\end{table}

\subsection{Formulation of the model}

In its original form, the $SIR$ model assumes perfect mixing and therefore
spatial homogeneity.
Therefore, spatial gradients are absent and $\kappa=0$.
The basic equations, with $\kappa\neq0$, are
\begin{equation}
{\partial S\over\partial t}=-\lambda SI+\gamma' R,
\label{dSdt}
\end{equation}
\begin{equation}
{\partial I\over\partial t}=\lambda SI-\mu I+\gamma R+\kappa\nabla^2 I,
\label{dIdt}
\end{equation}
\begin{equation}
{\partial R\over\partial t}=\mu I-(\gamma+\gamma') R,
\end{equation}
where $\lambda$ is the reproduction rate, $\mu$ is the rate of recovery,
while $\gamma$ and $\gamma'$ characterize the rates of reinfection
either directly via $I$ or by producing susceptible first, respectively.
The latter case ($\gamma=0$ with $\gamma'\neq0$) is also known as the
SIRS model.
Modeling reinfections through $\gamma'\neq0$ instead of $\gamma\neq0$ can
result in a slight reduction of $\bra{I}$, especially when $\mu$ is large.

Note that the model preserves the total population, i.e.,
$S+I+R=\const\equiv S_0$ when $\kappa=0$ and
$\bra{S+I+R}=\const\equiv S_0$ when $\kappa\neq0$, where
angle brackets denote an average over the population.
Here, $S_0$ is the initial population.
Therefore, only two of the three equations need to be solved.

In essence, the standard version of the model with $\gamma=\gamma'=0$
describes the increase of cases based on the current number of susceptible
individuals.
Once this number begins to be depleted, it can only decrease, although
it can still increase in neighboring locations, where the number of
cases may still be smaller.
This leads to spatial spreading of the disease and thereby ultimately
to an increase in the total number of cases.
Thus, the $SIR$ model with spatial extend is capable of describing
the increase of cases and deaths.
It remains unclear, however, whether the spatial increase corresponds
to a complete or only partially space-filling increase in the number of
cases over the surface of the Earth.
Given that the current number of cases now reaches a significant fraction
of the total population on Earth,\footnote{As of December 2022, three
years after the start of the pandemic, the fraction of the cases worldwide
is 8.5\%.
In the US and in France, for example, it is 31\% and 60\%, respectively.}
it may indeed be plausible that soon every single individual on Earth
is and was susceptible to the disease.

We solve \Eqs{dSdt}{dIdt} in a two-dimensional Cartesian domain with
coordinates $\xx=(x,y)$ and periodic boundary conditions.
We characterize the domain size $L$ by the smallest wavenumber $k=2\pi/L$
that fits into the domain.
We use the {\sc Pencil Code}, a publicly available time stepping code for
solving partial differential equations on massively parallel computers
\cite{JOSS}.
Spatial derivatives are computed from a sixth-order finite difference
formula and a third order Runge--Kutta time stepping scheme is employed.
As in Ref.~\cite{Bra20}, we use $4096^2$ mesh points and run the model
for about 1200 time units.
(During that time interval, the periodicity of the domain did not yet
play any role, because the disease did not reach the boundary.)
The $SIR$ model is implemented in the current version, and also the
relevant input parameter files are publicly available \cite{DATA}.

\begin{figure*}\begin{center}
\includegraphics[width=\textwidth]{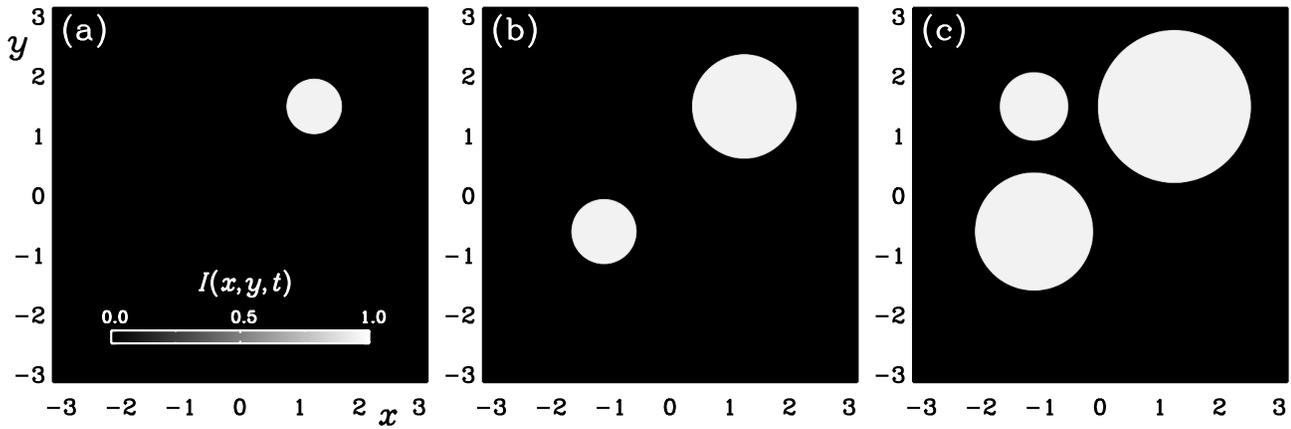}
\end{center}\caption[]{
$I(x,y,t)$ for $\mu=\gamma=\gamma'=0$ with $\kappa=10^{-6}$ and normalized
times (a) $\lambda t=300$, (b) $500$, and (c) $700$.
}\label{psav_rein_new}\end{figure*}

We define nondimensional space and time coordinates as $\tilde{\xx}=k\xx$
and $\tilde{t}=\lambda t$, respectively.
Furthermore, $\tilde{\mu}=\mu/\lambda$, $\tilde{\gamma}=\gamma/\lambda$,
$\tilde{\gamma}'=\gamma'/\lambda$, and $\tilde{\kappa}=\kappa k^2/\lambda$
are the only nondimensional input parameters that will be varied.
The population is normalized by $S_0$, so we can define $\tilde{S}=S/S_0$,
$\tilde{I}=I/S_0$, and $\tilde{R}=R/S_0$ as the fractional
(nondimensional) population densities.
We then have $\bra{\tilde{S}+\tilde{I}+\tilde{R}}=1$ at all times.
The tildes will from now on be dropped.
In practice, we keep $\lambda=1$ and adopt for the domain size $L=2\pi$,
so $k=1$.
For clarity, we often retain the factor $\lambda$ in front of the time
to remind the reader of the normalization.
Similarly, we often keep the normalizations of $\mu$, $\gamma$,
and $\gamma'$ and quote for the diffusivity the combination
$\kappa k^2/\lambda$ instead of just $\kappa$.

As initial condition, we assume $S=1$ and $I=0$, except for those
mesh points, where we initialize $I=I_1$ on one isolated mesh point
and $I=I_2$ on eight others.
We refer to them as ``hotspot''.

\subsection{Emergence of hotspots}

The main result of Ref.~\cite{Bra20} was that each hotspot, once it
reaches locally its saturation value, continues to grow only by spreading
on the periphery.
This peripheral growth is always quadratic.
However, once a new hotspot emerges, the coefficient in the growth
law increases.
\Figs{psav_rein_new}{psqrt_merge_new} provide an example of this.
At $\lambda t=300$, there is just one patch and the derivative of
$\bra{I}^{1/2}$ with respect to $\lambda t$ is about $6\times10^{-4}$.
When the next patch is established, the derivative becomes about
$8\times10^{-4}$, and after the third, the derivative becomes about
$10^{-3}$.
Thus, it seems that with each patch, the derivative increases by
about $2\times10^{-4}$.
However, there is an offset by about $4\times10^{-4}$ for the first one.
There is also a strong spike early on at $\lambda t\approx150$.
The existence of these two features suggests that there is an additional
contribution to the overall growth that is independent of the number
of patches.
These aspects are obviously not captured by the simple peripheral
growth model discussed in the introduction.
Interestingly, piecewise constant time derivatives have previously been
seen in a system where two populations compete against each other and
one of the two eventually disappears \cite{BM04}.
In that case, it was the area that decreased linearly with time and
there was no offset as in the present case.

\begin{figure}\begin{center}
\includegraphics[width=\columnwidth]{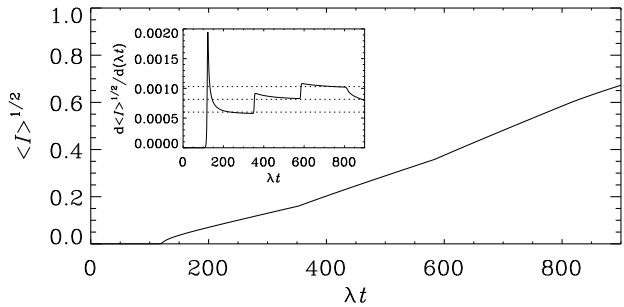}
\end{center}\caption[]{
Time series for a simulation with 3 hotspots of different strengths.
Note that $N^{1/2}\propto\bra{I}^{1/2}$ grows in a piecewise linear
fashion with time $t$.
The inset shows the normalized derivative $\dd\bra{I}^{1/2}/\dd(\lambda t)$.
The horizontal dotted lines mark the values $6\times10^{-4}$,
$8\times10^{-4}$, and $10^{-3}$.
}\label{psqrt_merge_new}\end{figure}

\subsection{Merging of patches}

\begin{figure*}\begin{center}
\includegraphics[width=\textwidth]{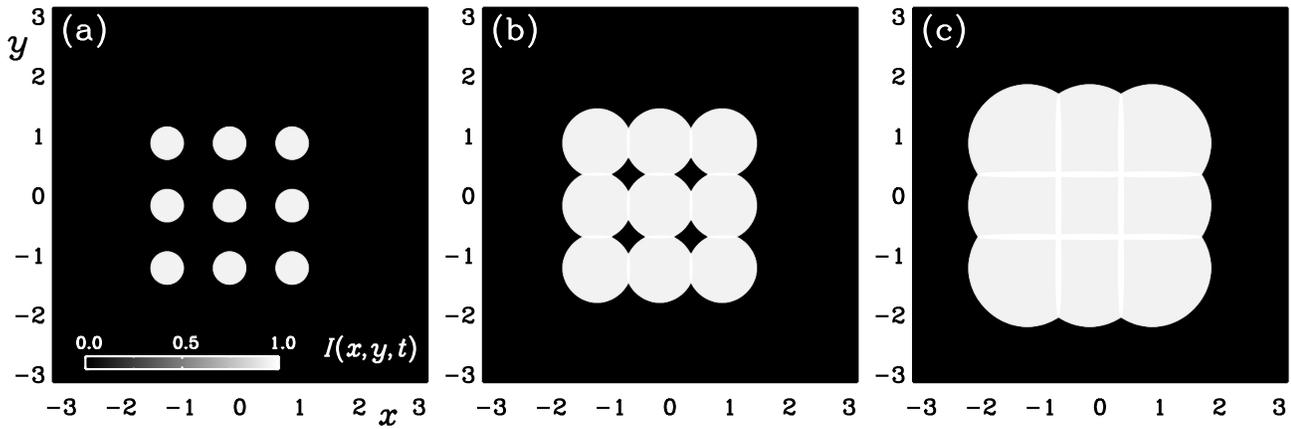}
\end{center}\caption[]{
Simulation with 9 hotspots that later merge and overlap.
The local distribution of $I(x,y,t)$ is shown in the $xy$ plane
for three values of $t$ ($\lambda t=150$, 300, and 500).
The length of the circumference determines the speed of growth.
When several hotspots merge, the circumference shortens and
the growth slows down.
Here, $\mu=\gamma=\gamma'=0$ and $\kappa k^2/\lambda=10^{-6}$.
}\label{psav_merge}\end{figure*}

\begin{figure}\begin{center}
\includegraphics[width=.99\columnwidth]{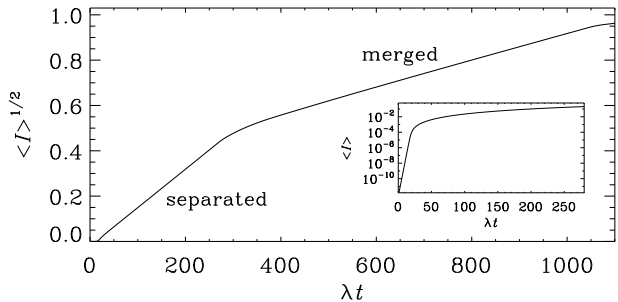}
\end{center}\caption[]{
Time series for simulation with 9 hotspots that later overlap.
Note that $N^{1/2}\propto\bra{I}^{1/2}$ grows linearly
with time $t$, which shows that $N\propto t^2$.
The inset shows the early exponential growth phase.
}\label{psqrt_merge}\end{figure}

The main purpose of the present study is to explain what happens when
different hotspots begin to merge at some moment.
\Fig{psav_merge} shows an example with initially nine separated hotspots.
In this model, $\mu=\gamma=\gamma'=0$, so there is no recovery and no
reinfections.
As before, the diffusivity is $\kappa k^2/\lambda=10^{-6}$.
By the time $\lambda t=300$, all patches have begun to merge, so the
total length of the periphery has decreased, and therefore also
the slope of growth has decreased; see \Fig{psqrt_merge}.
Quantitatively, the slope decreased from about $1.7\times10^{-3}$ to
about $6\times10^{-4}$, which is the same value that was found for a
single patch.

We argue that the phenomenon of merging models qualitatively the decrease
of the slope in \Fig{p3extn} at $t\approx100\days$ after January 20, 2020.
This implies that from that moment onward (beginning of May 2020),
the disease has begun to affect the entire world and the speed of
spreading was limited only by the containment efforts that took
place everywhere.

What the original $SIR$ model was not taking into account is the concept
of reinfection ($\gamma\neq0$ or $\gamma'\neq0$), i.e., the fact that
infected people can, after a certain period of time, be infected again.
This means that we must account for a term that describes the decrease of
reinfected individuals, which leads to a source in the number of people
that can be infected.
The purpose of the following is to explore in more detail the effect of
the sustainment of cases by the phenomenon of reinfection.

\subsection{Models with reinfection}

\begin{figure*}\begin{center}
\includegraphics[width=.99\textwidth]{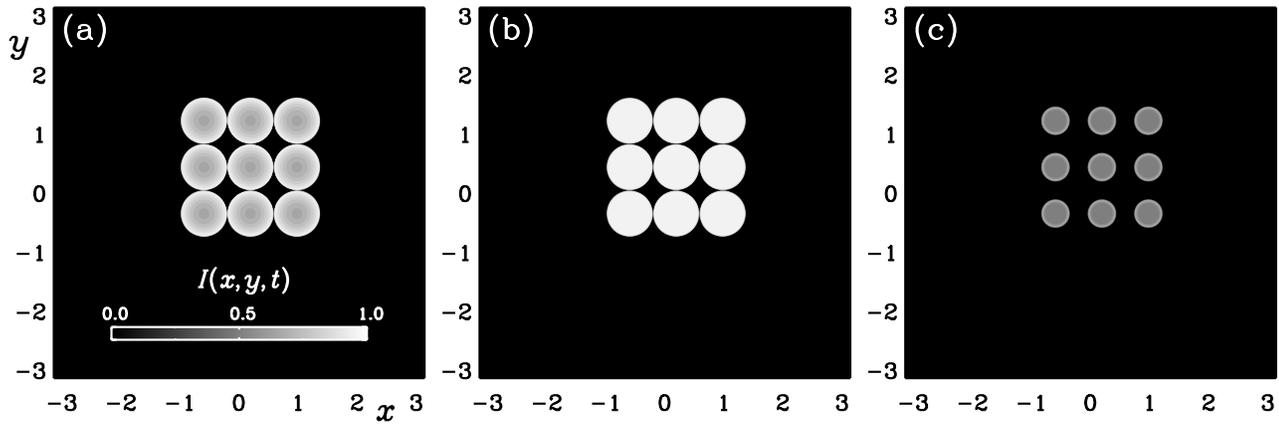}
\end{center}\caption[]{
Similar to \Fig{psav_merge}, but for models with
(a) $\mu/\lambda=5\times10^{-3}$ and $\gamma=0$,
(b) $\mu/\lambda=5\times10^{-3}$ and $\gamma/\lambda=0.1$, and
(c) $\mu/\lambda=0.1$ and $\gamma/\lambda=0.1$, all at $\lambda t=500$.
They illustrate that reinfections ($\gamma/\lambda=0.1$) lead to an increase (b),
but that increase diminishes significantly when the rate of recovery is
increased from $\mu/\lambda=5\times10^{-3}$ to 0.1 (c).
}\label{psav_rein}\end{figure*}

Next, we study models where the effect of reinfections is included,
i.e., $\gamma\neq0$.
In \Figs{psav_rein}{pcomp_rein}, we consider models with different
values of $\gamma$ and compare with cases where $\gamma'\neq0$.
We only study cases with $\mu\neq0$, because otherwise there are no
recoveries ($R=0$) and hence also no reinfections are possible.
We begin by considering here a relatively small value of
$\mu/\lambda=5\times10^{-3}$ and then also take a larger one of 0.1;
see Figure~10 of Ref.~\cite{Bra20} for other experiments with those
values of $\mu$.
When $\mu$ is small, there is a slow decline in the cores of each of patch;
see \Fig{psav_rein}(a).
When there are reinfections ($\gamma\neq0$ or $\gamma'\neq0$), the
cores are being prevented from depleting all the way to zero and
thus level off at a finite value of a about 0.95 for $\gamma/\lambda=0.1$;
see \Fig{psav_rein}(b).

We also studied a more extreme case where we $\mu/\lambda=0.1$.
In that case, even for $\gamma/\lambda=0.1$, the level of infections
remains at a residual level of about $0.55$.
This continued growth models the behavior seen at later times in
\Fig{p3extn}.

\begin{figure}\begin{center}
\includegraphics[width=.99\columnwidth]{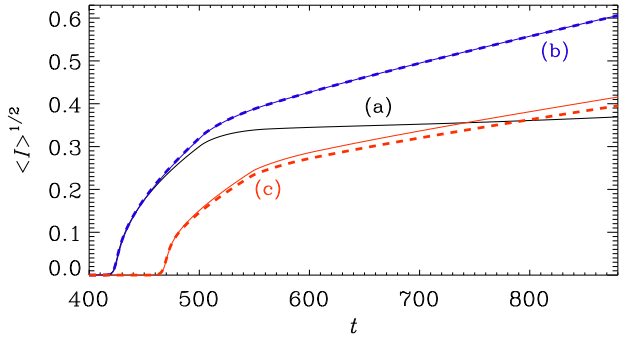}
\end{center}\caption[]{
Similar to \Fig{psqrt_merge}, but for the cases (a)--(c) of
\Fig{psav_rein}.
The fat dashed red (b) and blue (c) lines denote cases where $\gamma=0$
and $\gamma'/\lambda=0.1$ has been chosen.
They show that the choice of the specific reinfection model has only
a small effect and leads to a mild decrease of $\bra{I}^{1/2}$ when
$\mu/\lambda=0.1$ (c).
}\label{pcomp_rein}\end{figure}

\Fig{pcomp_rein} shows that the two models (b) and (c) with
$\gamma/\lambda=0.1$ (or $\gamma'/\lambda=0.1$ for the dashed lines) have nearly
the same spreading speed.
This has to do with the fact that the spreading speed is primarily
determined by the diffusivity, as will be addressed next.
The larger rate of recovery in (c) is responsible for the downward
shift compared with (b).

\subsection{Diffusion determines expansion speed}

\begin{figure*}\begin{center}
\includegraphics[width=.99\textwidth]{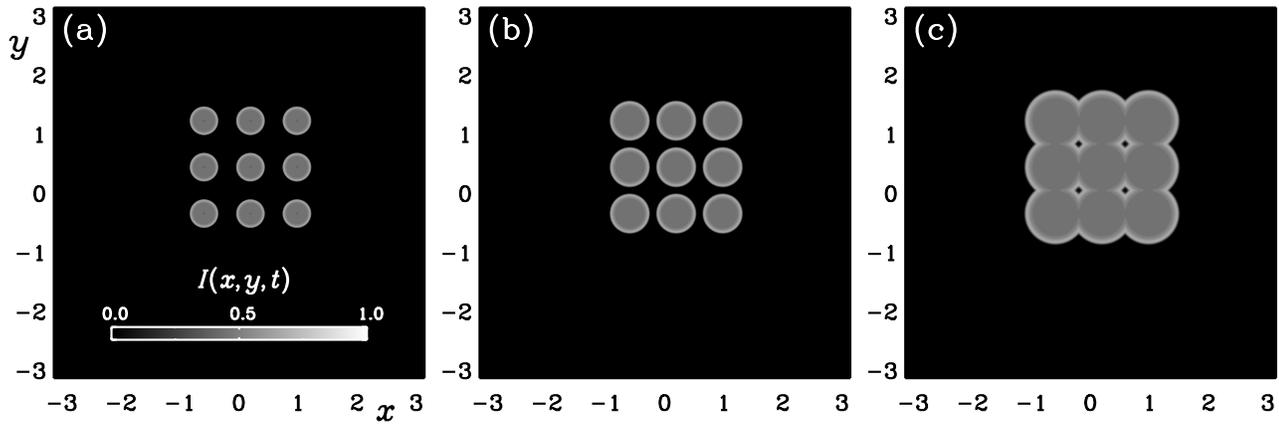}
\end{center}\caption[]{
Similar to \Fig{psav_merge}, but for models with $\mu/\lambda=0.1$,
$\gamma=0$, and $\gamma'/\lambda=0.1$ with (a) $\kappa k^2/\lambda=10^{-6}$,
(b) $2\times10^{-6}$, and (c) $5\times10^{-6}$, all at $\lambda t=500$.
Note that a larger diffusivity leads to larger patches in a fixed
amount of time.
For the largest diffusivity $\kappa k^2/\lambda=5\times10^{-6}$ (c),
the patches do already overlap at $\lambda t=500$.
}\label{psav_rein_diff}\end{figure*}

\begin{figure}\begin{center}
\includegraphics[width=.99\columnwidth]{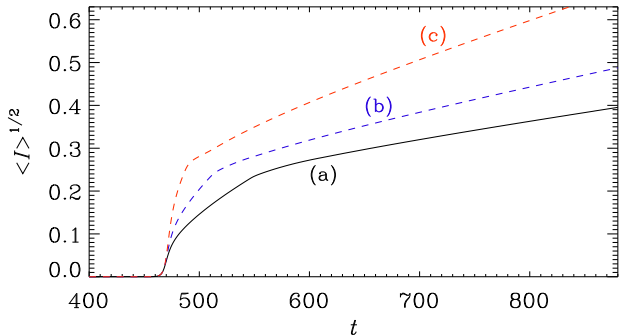}
\end{center}\caption[]{
Similar to \Fig{psqrt_merge}, but for the cases (a)--(c) of
\Fig{psav_rein_diff}.
Note that the slopes increase with increasing values of $\kappa$,
but decrease again once the patches begin to merge, which is the
case at $\lambda t\approx490$ (a), 512 (b), and 550 (c) for the
three values of $\kappa$.
Note also that panel (a) corresponds to \Fig{pcomp_rein}(c),
where reinfection is modeled by $\gamma/\lambda=0.1$ instead of
$\gamma'/\lambda=0.1$ in the present case.
}\label{pcomp_rein_diff}\end{figure}

As already emphasized in Ref.~\cite{Bra20}, and as expected from the
theory of epidemic front propagation \cite{Noble74, Murray+86}, the
value of $\kappa$ determines the speed of expansion.
This is shown in \Fig{psav_rein_diff}, where we compare models with
finite rates of recovery ($\mu/\lambda=0.1$) and reinfection by the
SIRS model ($\gamma'/\lambda=0.1$), and three different diffusivities
($\kappa k^2/\lambda=10^{-6}$, $2\times10^{-6}$, and $5\times10^{-6}$),
all at $\lambda t=500$.
We clearly see that the speed increases with increasing values
of $\kappa$ and that the patches are correspondingly larger after
the same amount of time.
The corresponding time traces for those three values of $\kappa$ are
compared in \Fig{pcomp_rein_diff}.
Here we see that the slopes increase with increasing values of $\kappa$,
but decrease again once the patches begin to merge.

\Fig{psav_rein_diff}(a), where $\gamma'/\lambda=0.1$ and $\gamma=0$, also shows
that the choice of the specific reinfection model is not important for
the final result.
This can be seen by comparing with \Fig{pcomp_rein}(c), which is the
same model, except that here $\gamma'=0$ and $\gamma/\lambda=0.1$.
Note that the corresponding time traces were already compared in
\Fig{psav_rein}; see the solid and dashed blue lines for case (c).

\subsection{Decreases in the slope}

\begin{figure}\begin{center}
\includegraphics[width=.99\columnwidth]{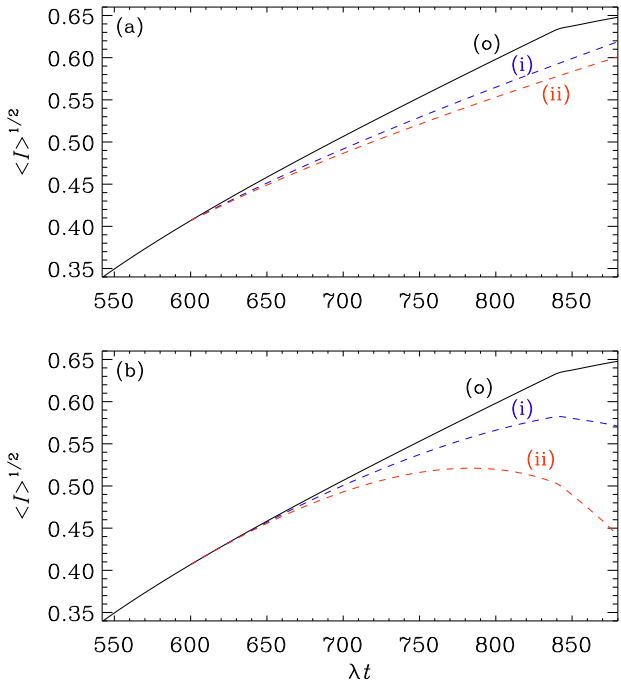}
\end{center}\caption[]{
Decreasing slopes from (a) decreasing values of $\kappa$ and
(b) decreasing reinfection rates $\gamma'$.
The reference run (o) is the same as (c)
in \Figs{psav_rein_diff}{pcomp_rein_diff} with $\mu/\lambda=\gamma'/\lambda=0.1$,
$\gamma=0$, and $\kappa k^2/\lambda=5\times10^{-6}$.
In both cases we restart at $\lambda t=600$.
In (a), we use (i) $\kappa k^2/\lambda=2\times10^{-6}$ and
(ii) $10^{-6}$, while in (b) we use \Eqs{gamma}{Theta} and
(i) $\lambda t_2=1000$ and (ii) $\lambda t_2=1200$.
}\label{pcomp_rein_diff_vary}\end{figure}

Given that the expansion speed depends on the value of $\kappa$, one
must expect that a time-dependent decrease of $\kappa$ should lead
to a decrease in the spreading speed.
This is shown in \Fig{pcomp_rein_diff_vary}(a), where we used the model
of \Fig{psav_rein}(c) and restarted it at $\lambda t=600$ with smaller
diffusivities.

Likewise, a decrease in the reinfection rate should also lead
to a decrease in the speed of spreading.
It turns out, however, that a sudden decrease in the reinfection rate,
for example from $\gamma'/\lambda=0.1$ to 0.05, has immediately a rather
noticeable effect on $\bra{I}^{1/2}$.
It is therefore desirable to let $\gamma'$ decrease in a smooth fashion.
Here we have chosen a modulation of the form
\begin{equation}
\gamma'=\gamma_0'\,\Theta(t;\;t_1,t_2),
\label{gamma}
\end{equation}
where
\begin{eqnarray}
\label{Theta}
\Theta(t)=
\max\left\{0,\;1-
\left[\frac{\max(0,\;t-t_1)}{t_2-t_1}\right]^2\right\}^2
\end{eqnarray}
is a function that goes smoothly from unity to zero between $t=t_1$ and
$t=t_2$.
The additional arguments $t_1$ and $t_2$ have here been suppressed
for brevity.

In \Fig{pcomp_rein_diff_vary}(b), we show cases with $\lambda t_2=1200$
(i) and $\lambda t_2=1000$ (ii), restarting again at $\lambda t_1=600$.
The results are promising and can serve to explain the decreases in the
slope seen in \Fig{p3extn}.

\section{Conclusions}

The present work has shown that the number of infected people will not
increase exponentially, as expected for a well mixed model without spatial
extend, but that it can increase instead quadratically and can be both
slower, if the local number of infected people is already exhausted,
and faster, if the number of susceptible people can still increase in
neighboring locations.

Obviously, the local number of cases cannot increase indefinitely, but
it can increase owing to the fact that the people in neighboring
locations can be infected and infected people can even be reinfected.
In the present work, we have studied in more detail the effect of
reinfections, which is especially important in cases when most of the
population has already been infected.
It turns out that a decrease in the total number of infections worldwide
can be explained by the merging of originally separated spreading centers.
In the case of SARS-CoV2, this happened at more or less the same time
(around $t\approx50\days$ in \Fig{p3extn}) for all the different spreading
centers on the Earth.

Subsequent variations in the number of cases and deaths can be explained
by variations in the reinfection rate.
This has been demonstrated by decreasing $\gamma'$ after a certain time,
and it led to a decrease in the spreading speed.
Similar results can also be reproduced by decreasing the diffusivity
at some time.
This would model a tightening of the control interventions and containment
regulations, but this is unlikely to explain the actual decrease in the
slope seen in \Fig{p3extn}.
Instead, a gradual decrease in the reinfection rate appears to be the
more plausible phenomenon causing the monotonic decrease in the slope
see in \Fig{p3extn}.

It is worth reflecting again on the meaning of patches.
It is not evident that the spreading of the disease can really be
described through patches.
However, the ability of our model in explaining quadratic growth is
rather generic and we may therefore be tempted to search more thoroughly
for an appropriate interpretation.

In this context, it is important to emphasize that quadratic growth
is not just an unspecified realization of governmental containment
efforts and control interventions of the disease, as was originally
speculated \cite{Maier+Brockmann20, Barzon+21}.
Instead, containment efforts may really mean that much of the population
was really excluded from the original spreading centers, and that the
disease is also so contagious that perfect containment was never possible,
so that there was always some leakage out of the patches or hotspots.
Within each patch, on the other hand, the level of infections is always
essentially saturated, which also explains why the early {\em exponential}
growth of the disease was so short.
This is also seen in our present simulations see; see the inset
of \Fig{psqrt_merge}.

In conclusion, our findings and interpretations of quadratic growth
are not so much a statement that we can predict a disease outcome, and
certainly not easily at the national level, but it is rather a way of
charactering the nature of SARS-CoV2 as an extremely contagious disease
that will easily spread locally to the maximum possible level and can then
be characterized as peripheral diffusive growth for each of the patches.
By now, SARS-CoV2 has almost affected the entire population, and yet,
the case numbers keep slowly increasing; see \Tab{TAB}.
Within our $SIR$ model with spatial extent, this can be described by
including reinfections in our equations.
The level of reinfections can easily be fluctuating because of
seasonal and other effects, which explains the long period of
growth with piecewise different (and mostly decreasing) slopes.

\section*{Acknowledgments}

I thank the two anonymous referees for useful comments that have led
to additional simulations and discussions in the paper.
This work was supported in part through the Swedish Research Council,
grant 2019-04234.
Nordita is sponsored by Nordforsk.
We acknowledge the allocation of computing resources provided by the
Swedish National Infrastructure for Computing (SNIC)
at the PDC Center for High Performance Computing Stockholm
and the National Supercomputer Centre (NSC) at Link\"oping.
\vspace{2mm}

\section*{Code and data availability statement}

The source code used for the simulations of this study,
the {\sc Pencil Code} \cite{JOSS}, is freely available on
\url{https://github.com/pencil-code/}, with datasets on Zenodo,
doi:10.5281/zenodo.4016941.

\section*{ORCID}

Axel Brandenburg:\\ \url{http://orcid.org/0000-0002-7304-021X}

\section*{References}

\bibliography{ref}
\end{document}